\documentclass{article}
\usepackage{epsf}
\usepackage{emulateapj}
\input{psfig}
\usepackage{mathptm}
\def\spose#1{\hbox to 0pt{#1\hss}}
\def\lta{\mathrel{\spose{\lower 3pt\hbox{$\mathchar"218$}}
     \raise 2.0pt\hbox{$\mathchar"13C$}}}
\def\gta{\mathrel{\spose{\lower 3pt\hbox{$\mathchar"218$}}
     \raise 2.0pt\hbox{$\mathchar"13E$}}} 
\def\etal{et~al} 
\def\ergcm2s{${\rm erg~cm^{-2} s^{-1}}$}
\def\ergscm2s{${\rm erg~cm^{-2} s^{-1}}$}

\def\cm2{${\rm cm^{-2}}$}
\def\ergs{${\rm erg~s^{-1}}$}

\def\sax1808{SAX~J1808.4-3658}
\def\msol   {{M$_{\odot}$}}

\begin{document}

\lefthead{GARCIA et al}
\righthead{Chandra Observations of Black Holes}
\slugcomment{Submitted to ApJ Letters Dec 20, 2000}

\title{New Evidence for Black Hole Event Horizons from Chandra}

\author{Michael R. Garcia,\altaffilmark{1} Jeffrey
E. McClintock,\altaffilmark{1} Ramesh Narayan,\altaffilmark{1} Paul
Callanan,\altaffilmark{2} \& Stephen S. Murray\altaffilmark{1}}

\altaffiltext{1}{Harvard-Smithsonian Center for Astrophysics, 60
Garden Street, Cambridge, MA 02138, USA.\\ mgarcia@cfa.harvard.edu,
jmcclintock@cfa.harvard.edu, rnarayan@cfa.harvard.edu,
smurray@cfa.harvard.edu} \altaffiltext{2}{University College, Cork,
Ireland, paulc@ucc.ie}

\begin{abstract}

Previously we claimed that Black Hole X-ray Novae (BHXN) in
quiescence are much less luminous than equivalent Neutron Star X-ray Novae
(NSXN).  This claim was based on the quiescent detection of a single
short period BHXN (A0620--00, P$_{orb}$=7.8 hrs) and two longer period
BHXN (GRO~J1655--40, P$_{orb}$=62.9 hrs; V404~Cyg, P$_{orb}$=155.3
hrs), along with sensitive upper limits.  We announce the detection of
two more short period BHXN (GRO J0422+32, P$_{orb}$=5.1~hrs;
GS~2000+25, P$_{orb}$=8.3~hrs), an upper limit for a third which is
improved by two orders of magnitude (4U 1543--47, P$_{orb}$=27.0~hrs)
and a new, much lower quiescent measurement of GRO~J1655--40.  Taken
together, these new Chandra measurements confirm that the quiescent
X-ray luminosities of BHXN are significantly lower than those of NSXN.
We argue that this provides strong evidence for the existence of event
horizons in BHXN.

\end{abstract}


\section{Introduction}

The very low, but non-zero, quiescent X-ray luminosity of the BHXN
A0620--00 is difficult to understand in the context of standard
viscous accretion disk theory (McClintock, Horne \& Remillard 1995),
given the continued mass transfer from the companion evidenced by an
optically bright disk.  The observations are explained by an
advection-dominated accretion flow model, or ADAF (Narayan, McClintock
\& Yi 1996).

In an ADAF (see Narayan, Mahadevan \& Quataert 1998 and Kato, Fukue \&
Mineshige 1998 for reviews), the energy released by accretion is
stored as heat in a radiatively inefficient flow.  Should the
accreting object be a black hole this energy would be lost from view
once it crosses the event horizon, but if the object has a solid
surface the energy would be released upon impact with that surface and
radiated to infinity.  Thus, for the same mass accretion rate, a black
hole would be significantly less luminous than a compact star with a surface
(Narayan \& Yi 1995).  Such a comparison of otherwise similar systems
is a promising method for proving the reality of event horizons
(Narayan, Garcia \& McClintock 1997a, hereafter NGM).

X-ray novae (XN) containing black hole (BH) and neutron star (NS)
primaries are believed to be similar in many respects.  Most
importantly, the mass transfer rate from the secondary, measured in
Eddington units of the primary, is believed to be comparable in BHXN
and NSXN of similar orbital periods (Menou \etal\/ 1999, hereafter
M99).  NGM, Garcia et al. 1998 (hereafter G98) and M99 showed that
quiescent BHXN are much less luminous in X-rays than quiescent NSXN of
similar orbital periods, and argued that this provides direct evidence
that BHXN are able to ``hide'' their accretion energy behind an event
horizon.  However, their claims were based on a very small sample in
which only one detected BHXN (A0620--00, $P_{orb}=7.8$ hrs) had an orbital
period similar to those of the comparison NSXN.

We report here new observations of four BHXN as part of the Chandra
AO1 HRC Team GTO program and an additional two BHXN as part of a
complementary AO1 GO program.

\section{Observations and Analysis}

The basic parameters of the Chandra observations described here are
presented in Table~1.  Data were analyzed with a combination of the
CXC CIAO V1.1 (CXC 2000a), HEASARC XSPEC V11.0 (Arnaud \&
Dorman 2000), and software written by Alexey Vikhlinin (Vikhlinin
\etal\/ 1998).  Detected fluxes and counting rates were measured over
a 0.3--7.0 keV energy band for ACIS-S observations, and 0.5--7.0 keV
for ACIS-I observations, in order to reduce the instrumental
background.  The observed on-axis PSF is such that 95\% of the source
flux is contained in the $1.5''$ radius source extraction circle (van
Speybroeck \etal\/ 1997; CXC 2000b).  We observed occasional
intervals of enhanced background during the observation of A0620--00
(see Plucinsky \& Virani 2000), and therefore rejected 6.7~ks of this
observation.  The total accumulated background within the source
extraction circle is $<1$~count per source for all the observations.

We defer a full discussion of the Chandra spectra of quiescent BHXN to
a second paper (McClintock \etal\/ 2001).  Here we merely note that
all the Chandra spectra are consistent with a single power-law with
$\alpha \sim 2$ and absorption consistent with the optical extinction.
We assume this in estimating the luminosities quoted here (unless
explicitly stated otherwise).

\begin{table*}
\caption{Chandra Observations of Quiescent BHXN (Detected Fluxes and Limits)}
\begin{center}
\begin{tabular}{llrcccccl} \hline \hline
\\

System & Instrument & Counts & Exposure & $f_x$ & L$_{\rm X}$   & $D$& m$_1$&log ${\rm N_H}$ \\
	&		&		& ks	&ergs~s$^{-1}$cm$^{-2}$ & ergs~s$^{-1}$ & kpc &\msol& cm$^{-2}$ \\ \hline

GRO J0422+32		& ACIS-I & 16	& 18.8 & $6.6\times 10^{-15}$&$7.6\times 10^{30}$&2.6&12&21.3\\
A0620--00 (XN Mon 75) 	& ACIS-S & 123  & 38.3 & $1.8\times 10^{-14}$&$2.8\times 10^{30}$&1.0&6.1&21.3\\
GS~2000+25 (XN Vul 88) 	& ACIS-I & 5  	& 21.5 & $2.4\times 10^{-15}$&$2.4\times 10^{30}$&2.7&8.5&21.9\\
4U1543--47		& ACIS-I & $<5$ &~9.9 &$<4.2\times 10^{-15}$&$<3.0\times 10^{31}$&6.1$^a$&6$^a$&21.6$^b$\\
GRO J1655--40          	& ACIS-S & 65  	& 42.5 & $1.2\times 10^{-14}$&$2.4\times 10^{31}$&3.2&7&21.8\\   
V404 Cyg (GS2023+338) 	& ACIS-S & 1655 & 10.3 & $1.5\times 10^{-12}$&$4.9\times~10^{33}$&3.5&12&22.0\\\hline
\end{tabular}
\end{center}
Note: Values of ${\rm N_H}$ are taken from G98 and
NGM;  distances and primary masses are from M99, unless explicitly mentioned.
Detected ACIS-S fluxes are over a 0.3--7.0~keV band; ACIS-I fluxes are
over a 0.5--7.0~keV band; and emitted luminosities are over a
0.5--10.0~keV band.\hfill\\
$^a$Orosz \etal\/ 2001. \hfill\\
$^b{\rm N_H}$ from
Greiner, Predehl \& Harmon (1994) and van der Woerd, White \& Kahn
(1989).

\end{table*}

{\sl GRO J0422+32.---}The GRO J0422+32 region was observed with the
Chandra Observatory (Weisskopf \& O'Dell 1997) ACIS-I CCD array
(Garmire \etal\/ 1992) for 18.8~ks on 2000 Oct 10.  The CCD array
recorded 16~photons at a position consistent with the optical position
of GRO J0422+32, which constitutes a clear detection.  Assuming a
distance of 2.6~kpc (M99) and X-ray absorption corresponding to the
optical extinction of ${\rm A_V} = 1.2$ (Filippenko, Matheson, \& Ho
1995; Predehl \& Schmitt 1995), we derive an emitted luminosity of
$7.6 \times 10^{30}$\ergs\/ (0.5--10.0~keV).  Luminosities, distances,
primary masses, and log ${\rm N_H}$ for all sources are listed in
Table~1.

{\sl A0620--00.---}The A0620--00 region was observed for 45~ks on 2000
Feb 29 and 38.3~ks of low background data was obtained.  
A0620--00 was placed on the ACIS-S3 CCD (Garmire \etal\/ 1992)
in order to maximize the counting rate from the possibly soft X-ray
spectrum (Narayan, Barret, \& McClintock 1997b).  
The CCD recorded 123~photons at a
position consistent with the optical position of A0620--00.

{\sl GS~2000+25.---} The GS~2000+25 region was observed with the Chandra
ACIS-I detector on 1999~Nov~5 for a total of 21.5~ks.  The CCD array 
recorded 5~photons near the position of GS~2000+25, which is a weak
but significant detection.
In order to more accurately determine the position of this source, we
cross correlated the positions of all sources in the image with
$>$10~counts against the USNO-A2 catalog.  We found 4~matches within a
$2''$ search radius.  We then used the IRAF ctio.coords task to
determine the WCS for the X-ray image using these 4~stars.  The plate
solution has an RMS of $0.3''$, and shows that the ICRS position of
this weak source is RA=20:02:49.52, DEC=+25:14:10.34.  Our ability to
determine the centroid of this source is limited by the low number of
counts to $\sim 0.5''$.

We determined the position of the optical counterpart of GS~2000+25 in
a similar way, matching 8~USNO-A2 stars against a K-band image of the
field we acquired at the LICK 3m with the Gemini camera.  This plate
solution has an RMS of $0.2''$, and shows the position of the optical
(IR) counterpart to be RA=20:02:49.55,  DEC=+25:14:10.94.  The offset
between the two positions is $0.7'' \pm 0.8''$, thus 
the position of the weak X-ray source is consistent with GS~2000+25.  

{\sl 4U 1543--47.---} A 9.9~ks ACIS-I observation of 4U~1543--47 took
place on 2000 July 26, and failed to detect even a single photon from
this source.  We compute a conservative upper limit to the luminosity
by assuming a that $<5$ photons were detected.  This luminosity (see
Table~1) is a factor of $\sim 100$ below the previous upper limit
(M99).

{\sl GRO J1655--40.---} A 42.5~ks ACIS-S observation of GRO J1655--40
took place on 2000 July 2, and detected 65 photons from this source.
We note that this corresponds to a luminosity a factor of $\sim 10$
below the previously measured quiescent luminosity (G98).  Given the
quiescent variations seen in V404 Cyg (Wagner \etal\/ 1994), this
could simply be indicative of typical quiescent variations.
Alternately, we note that the previous quiescent observations were in
between two large outbursts separated by $\sim 1$~year, and therefore
may not have been indicative of the true quiescent level.  The Chandra
observations reported here occurred $\sim 4$~years after the last
outburst, and therefore may accurately measure the quiescent level.

{\sl V404 Cyg.---} A 10.3~ks ACIS-S observation of V404 Cyg took place on
26 April 2000, and allowed detection of 1655 photons from this source.

\section{Discussion}

We have collected in Table 2 the available quiescent X-ray
luminosities and upper limits for BHXN and NSXN.  We have supplemented
our Chandra measurements with the recent quiescent detections of
\sax1808 (Stella \etal\/ 2000; Dotani, Asai \& Wijnands 2000) and
included previous data compiled by M99.  Figure~1 displays the
Eddington-scaled luminosities as a function of orbital period
$P_{orb}$.  For calculating $L_{Edd}$, we used the BH mass estimates
in Table 1 and assumed that all NSs have a mass of $1.4M_\odot$. 

With the Chandra detections of GRO~J0422+32 and GS~2000+25 reported
here, three short period BHXN have now been detected in quiescence.  A
greatly improved upper limit for a fourth short period system,
4U1543--47, is nearly comparable to these detections.  In addition, we
find a new, ten-times fainter quiescent luminosity for GRO~J1655--40,
which has a period intermediate between the NSXN and the long period
BHXN V404~Cyg.

The new data points significantly strengthen our earlier claim (NGM,
G98, M99) that BHXN have much lower (roughly a factor
of 100) quiescent luminosities than NSXN.  As explained in \S1, such a
difference in luminosity is natural if quiescent accretion proceeds
via a radiatively inefficient ADAF {\it and} if the primaries in BHXN
have event horizons.  Thus, the new data bolster the evidence for
event horizons in BHXN.

Our argument relies on the reasonable assumption that the Eddington
scaled mass accretion rate is similar in quiescent BHXN and NSXN.  At
the short orbital periods characteristic of the NSXN in our sample,
angular momentum loss through gravitational radiation is expected to
be the dominant mechanism driving mass transfer from the secondary.
The Eddington-scaled mass transfer rates are then likely to be roughly
similar in BHXN and NSXN of similar $P_{orb}$ (M99).  At the long
period of V404~Cyg and perhaps GRO~J1655--40, nuclear evolution is
expected to drive the mass transfer rate to substantially higher
values, and so these systems are less useful for our purposes.
Another reason for ignoring these systems is the fact that there are
no NSXN with similar $P_{orb}$ for comparison.

\vbox{
\centerline{
\psfig{figure=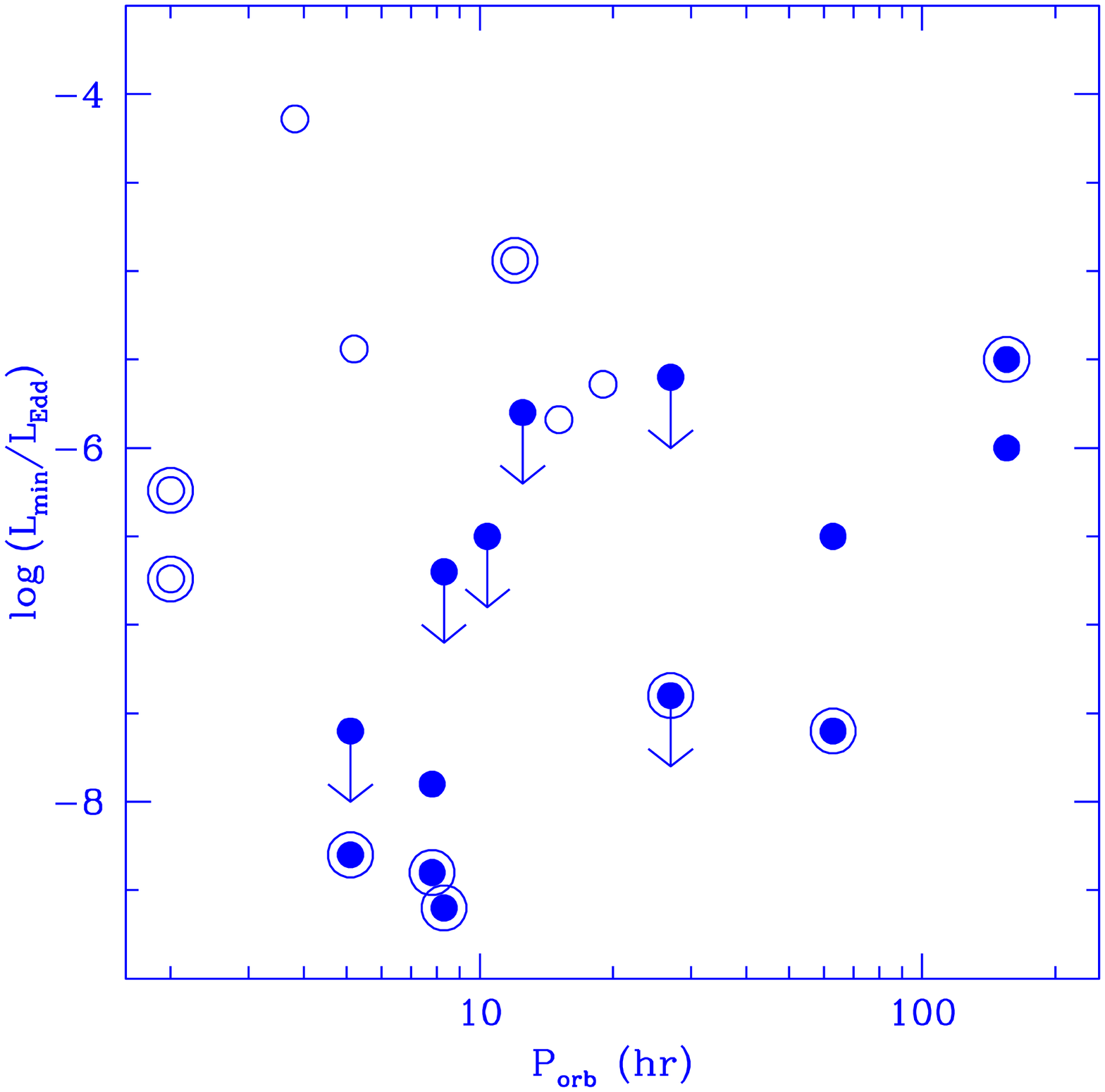,width=3.0in}}
\centerline{
\psfig{figure=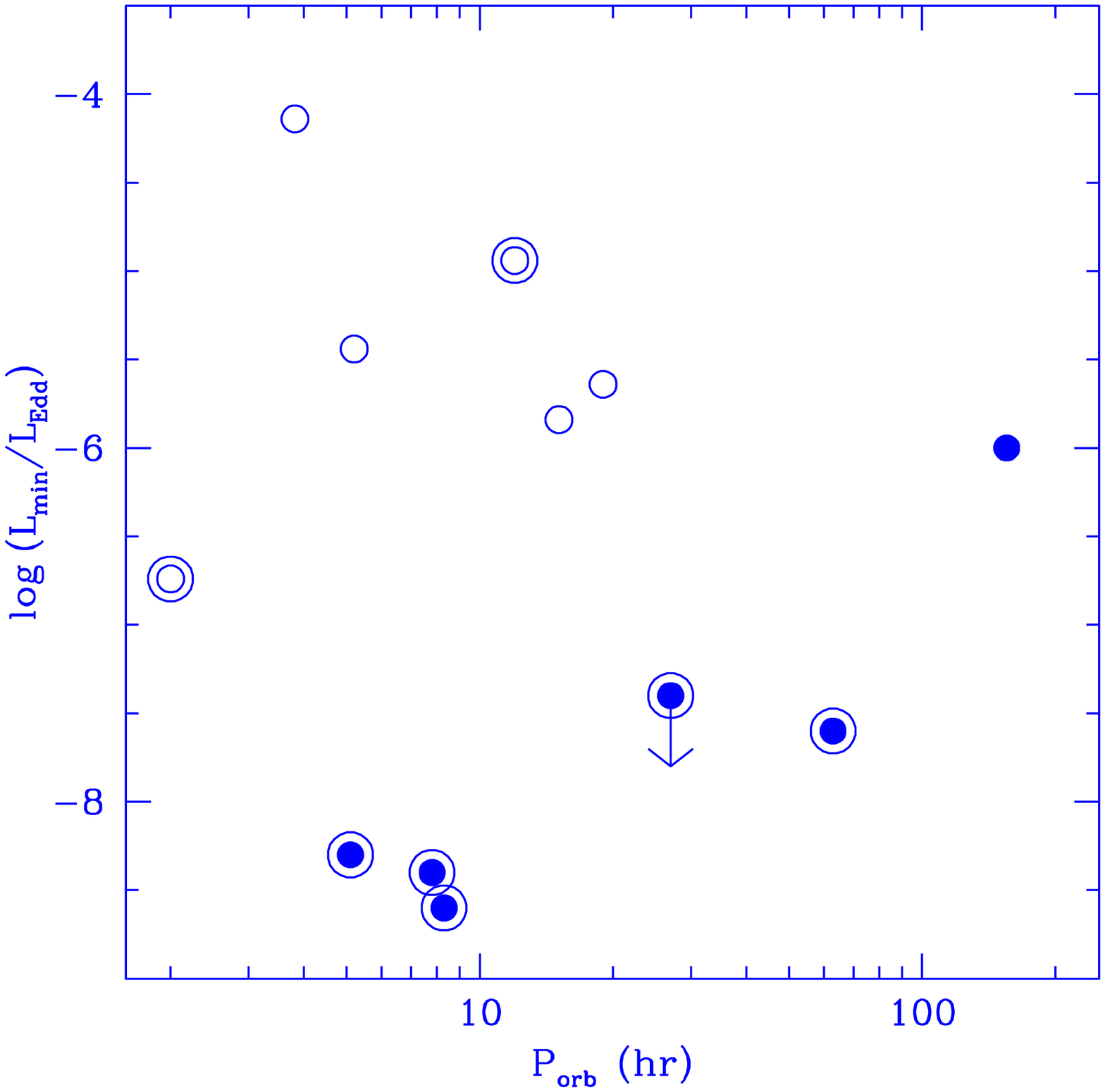,width=3.0in}}
\centerline{
\begin{minipage}[h]{3.5in}
{\small Figure 1 (Top): Quiescent luminosities of BHXN (filled
circles) and NSXN (open circles) from Table~2.  
Data points not included in M99 are
circled.  Multiple quiescent detections are included.  (Bottom): Shows
only the lowest quiescent detections or Chandra upper limits.}
\end{minipage}
}
}

We also assume that $f_*$, the fraction of matter transferred from the
secondary which actually reaches the central star, is the same in BHXN
and NSXN.  Out-flowing winds from an ADAF (Narayan \& Yi 1994, 1995,
Blandford and Begelman 1999, M99) tend to reduce $f_*$, but there is
no reason to expect winds to be stronger (by a factor of 100) in BHXN
than in NSXN.  In fact, a centrifugal propeller and/or radio pulsar
action could reduce $f_*$ in NSXN without affecting BHXN (NGM, M99).
Therefore, if at all, we expect $f_*$ to be lower in NSXN than in
BHXN, rather than the other way round.  This strengthens the case for
event horizons in BHXN.

In this connection, we note that \sax1808 is the least luminous of the
NSXN.  It also may have a time averaged mass transfer rate that is
lower than that typical of XN, perhaps due to the action of the radio
pulsar on the secondary (Chakrabarty \& Morgan 1998).  Given that
\sax1808 is the only example of a NSXN which shows permanent coherent
pulsations, it seems possible that it may also have an unusually
efficient ``propeller.''  In addition, this system may have already
evolved past the minimum orbital period and the orbit may now be
expanding (King 2000; Dotani \etal\/ 2000).  Despite all these
anomalies, this NSXN is $\sim 60$~times brighter than our three short
period BHXN (see Figure~2).

{\small
\begin{center}
Table 2 \\
Quiescent Luminosities \\
\medskip
\begin{tabular}{lcc} \hline \hline
\\

System & $P_{\rm orb}$ (hr) & $\log [L_{\rm min}]~({\rm erg~s^{-1}})$ \\ \\ (1) & (2) & (3) \\
\\ \hline \\

$\circ$ {\bf SAX J1808-365} & $2.0^a$  & {\bf 32.0$^b$, 31.5$^c$} \\
$\circ$ EXO 0748-676 	& 3.82           & 34.1    \\
$\circ$ 4U 2129+47 	& 5.2              & 32.8    \\
$\circ$ Cen X-4		& 15.1      & 32.4   \\
$\circ$ Aql X-1 	& 19       & 32.6   \\
$\circ$ {\bf H1608-52}  &{\bf 12}$^d$   & 33.3   \\
    
\\
\hline
\\

$\bullet$ {\bf GRO J0422+32} 	& 5.1 	&{\bf  30.9},$<31.6$   \\
$\bullet$ {\bf A0620--00}   	& 7.8 	&{\bf 30.5},30.8$^e$   \\
$\bullet$ {\bf GS~2000+25 } 	& 8.3 	&{\bf 30.4},$<32.2$    \\
$\bullet$ GS1124-683       	& 10.4 	&$< 32.4$     \\
$\bullet$ H1705-250 		& 12.5 	&$< 33.0$     \\
$\bullet$ {\bf 4U 1543--47}     	& 27.0 	&$<${\bf 31.5},$<33.3$	\\
$\bullet$ {\bf GRO J1655--40}   	& 62.9 	&{\bf 31.3},32.4   \\
$\bullet$ {\bf V404 Cyg} 	& 155.3 &{\bf 33.7},33.1$^f$	\\

\\
\hline

\end{tabular}
\begin{minipage}{3.0in}
NOTES --- References are as in M99, unless otherwise noted.  New and/or
 CHANDRA Measurements are {\bf in BOLD FACE}. 
(1) $\circ$ indicates a NS primary and $\bullet$ a BH
primary.  (2) Orbital periods. (3) Luminosities in
quiescence in the 0.5-10 keV band (corrected for the revised
 distances). \hfill\\
$^a$Chakrabarty and Morgan 1998. \hfill\\
$^b$Stella etal 2000; assuming D$=2.5$~kpc.\hfill\\
$^c$Dotani, Asai, and Wijnands 2000. \hfill\\
$^d$Wachter 2000 reports a new 12~hr  period,  intermediate between
 the previously reported 98.4~hr (Ritter  \& Kolb 1998) and 5~hr (Chen et al. 1998) periods.  \hfill\\
$^e$Recomputed for $\alpha = 2$ from NGM.\hfill\\
$^f$Kong 2000.\hfill\\
\end{minipage}
\end{center}
}

Campana \& Stella (2000) argue that it is not correct to compare only
X-ray luminosities, as we have done, but that we should include also
the quiescent non-stellar optical and UV luminosity.  This point is
not obvious since the origin of the optical/UV luminosity is presently
unclear.  Within the ADAF model, it depends on the poorly known radius
at which the inner edge of the accretion disk evaporates into the ADAF
(e.g., Narayan, McClintock \& Yi (1996); Narayan, Barret \& McClintock
1997b).  It also depends on the strength of winds from the ADAF
(Quataert \& Narayan 1999).  If the non-stellar optical/UV luminosity
originates in the outer accretion disk, or in the hot spot where the
mass transfer stream from the secondary impacts the disk, then the
fact that the optical/UV luminosities of BHXN and NSXN are similar
(Campana \& Stella 2000) provides observational confirmation that the
mass transfer rates in the two kinds of system are similar, as we have
assumed.  An 0.25 mag modulation in the far-UV flux of A0620--00 on an
orbital time scale suggests that the far-UV flux may be modulated with
the orbital phase (McClintock 2000), which indicates an origin in the
hot spot. 

Bildsten \& Rutledge (2000, but see Lasota 2000) suggest that the
X-rays in BHXN may be produced by a rotationally enhanced stellar
corona in the secondary, especially in the short period BHXN with the
lowest luminosities.  The detection of GRO~J0422+32 with a luminosity
of nearly $10^{31}~{\rm erg\,s^{-1}}$ rules out the coronal
interpretation for this system (Lasota 2000), and perhaps all the
other BHXN in our sample as well.  Also, the Chandra spectra of
A0620--00 and GRO~J1655--40 do not appear to be consistent with the
stellar corona interpretation (McClintock \etal\/ 2001).  If some (or
all) of the X-ray emission in BHXN were coronal, then the accretion
luminosities of the black holes would be even lower than our estimates
and the argument for event horizons would be further strengthened.

A critical element in our comparison of NSXN and BHXN quiescent
luminosities is the assumption that the quiescent X-ray luminosities
of NSXN result from accretion. Brown, Bildsten \& Rutledge (1998) have
suggested that the luminosity could be due to crustal heating of the
NS during outburst followed by cooling in quiescence.  The rapid
variability of the prototypical NSXN Cen X-4 (Campana \etal\/ 1997) is
hard to explain in a cooling model, and shows that at most $\sim 1/3$
of the quiescent luminosity of this source is due to crustal cooling
(M99).  In addition, Cen X-4 and Aql X-1 have substantial power-law
tails in their spectra (carrying about half the total luminosity), and
it is hard to explain this spectral component with a cooling model.
It thus seems reasonable to assume that accretion accounts for a
substantial fraction of the quiescent X-ray luminosity in most NSXN.
The optical variability of NSXN in quiescence (McClintock \& Remillard
2000; Jain \etal\/ 2000; Ilovaisky \& Chevalier 2000) provides ample
evidence that accretion continues during quiescence.

Regardless of the caveats mentioned above, ultimately the dramatic
difference in quiescent X-ray luminosities of BHXN and NSXN needs to
be explained.  Any explanation is likely to require a fundamental
difference in the nature of the primaries in BHXN and NSXN.  In our
view, any straightforward explanation, whether based on ADAFs or not,
will require postulating an event horizon in BHXN.

The Eddington scaled luminosities of some supermassive black holes
(SMBH) in galactic nuclei (e.g., Garcia \etal\/ 2000; Baganoff \etal\/
2000) are much smaller than those of the ``stellar mass'' black holes
discussed here.  Could this be evidence for the event horizon (Narayan
et al. 1998)?  Unfortunately, the mass accretion rates of SMBH are
difficult to determine observationally (DiMatteo 2000; DiMatteo,
Carilli, \& Fabian 2000; Quataert, Narayan, and Reid 1999), and one
cannot exclude the possibility that the low luminosities of SMBH are
simply the result of extremely low accretion rates.  In the case of
quiescent BHXN, we have the luxury of a control sample of NSXN.  By
comparing quiescent BHXN and NSXN with similar orbital periods, we
eliminate the uncertainty in the mass accretion rate, and thus obtain
more secure evidence for the event horizon.

\acknowledgements{This work was supported in part by NASA Contract
NAS8-39073 to the Chandra X-ray Center, contract NAS8-38248 to the
HRC Team, and NSF grant AST 9820686 to RN. 
We thank A. Kong for helpful comments.}

\end{document}